\documentclass[runningheads]{llncs}
\usepackage[T1]{fontenc}
\usepackage{graphicx}
\usepackage{hyperref}

\usepackage{color}   
\usepackage[misc]{ifsym}
\usepackage{makecell}
\usepackage{multirow}
\usepackage{subcaption}
\usepackage{amsmath}

\begin{document}

\title{Interpretable Prediction of Pulmonary Hypertension in Newborns using Echocardiograms}
%OPTIONS:
%An Ensemble Approach to Predict Pulmonary Hypertension in Newborns using Cardiac Ultrasounds with a focus on Interpretability.
% A fully automated diagnosis of Pulmonary Hypertension using pediatric echocardiograms with a focus on interpretability and robustness.
%Deep Learning to Predict Pulmonary Hypertension using Pediatric Echocardiograms
\titlerunning{Interpretable Prediction of Pulmonary Hypertension in Newborns}

%\author{Anonymous Author(s)}
 \author{Hanna Ragnarsdottir\inst{1} \and Laura Manduchi\inst{1}(\Letter) \and Holger Michel\inst{2} \and Fabian Laumer\inst{1} \and Sven Wellmann\inst{2} \and Ece Ozkan\inst{1}(\Letter) \and Julia Vogt\inst{1}}
\authorrunning{Author et al.}
\authorrunning{H. Ragnarsdottir et al.}
%\institute{Anonymous Institute(s)}
\institute{Department of Computer Science, ETH Zürich\\
\email{\{laura.manduchi,ece.oezkanelsen\}@inf.ethz.ch}
\and
Department of Neonatology, \\
University Children’s Hospital Regensburg (KUNO), Germany}

\maketitle              

\begin{abstract}
Pulmonary hypertension (PH) in newborns and infants is a complex condition associated with several pulmonary, cardiac, and systemic diseases contributing to morbidity and mortality. 
Therefore, accurate and early detection of PH is crucial for successful management.  
Using echocardiography, the primary diagnostic tool in pediatrics, human assessment is both time-consuming and expertise-demanding, raising the need for an automated approach.
In this work, we present an interpretable multi-view video-based deep learning approach to predict PH for a cohort of 194 newborns using echocardiograms. 
We use spatio-temporal convolutional architectures for the prediction of PH from each view, and aggregate the predictions of the different views using majority voting. 
To the best of our knowledge, this is the first work for an automated assessment of PH in newborns using echocardiograms. 
Our results show a mean F1-score of 0.84 for severity prediction and 0.92 for binary detection using 10-fold cross-validation. 
We complement our predictions with saliency maps and show that the learned model focuses on clinically relevant cardiac structures, motivating its usage in clinical practice.

\end{abstract}

\section{Introduction}
Pulmonary hypertension (PH) is a complex condition that can affect newborns and children as well as adults and is formally defined as an increased mean pulmonary artery pressure (PAP) at rest \cite{HANSMANN20172551}. % , specifically $PAP > 25$ mmHg. 
The level of PAP in newborns is frequently high and it is expected to decrease after birth to reach a level comparable to healthy adult values \cite{deBoode2018}. 
When the normal cardiopulmonary transition fails to occur, the newborns are affected by persistent pulmonary hypertension of the newborn (PPHN), which is associated with bronchopulmonary dysplasia in older premature infants and various chronic pulmonary, cardiac, and systemic diseases for newborns at term contributing to morbidity and mortality \cite{EL-Khuffash2014,HANSMANN20172551,Steinhorn2010NeonatalPH}.

Pulmonary hypertension is diagnosed primarily with echocardiography, which % The gold standard for PH diagnosis is right heart catheterisation (RHC). However, due to its invasive nature of this costly procedure, and the resulting high risk of related complications, especially in this age group \cite{Rosenkranz2015}, RHC is not a screening procedure and is often only used to validate the previous PH diagnosis \cite{Ni2019}. Echocardiography, on the other hand, 
is recommended as a non-invasive screening tool for PH using its two modes: 2D echocardiography videos (ECHOs) and Doppler echocardiograms \cite{Lang2015RecommendationsFC,Ni2019}. It is one of the most common and growing diagnostic tools due to its low-cost, portable, and non-invasive technology, which makes it an ideal choice for pediatrics \cite{Lang2015RecommendationsFC}. %  among others.
Screening of PH typically involves estimating PAP with Doppler echocardiography, however, the measurements may frequently be inaccurate, thus, is not the ultimate predictive tool to assess and manage PH \cite{Fisher2009}.
Since elevated PAP can result in abnormalities in the shape and structure of the heart, subjective evaluation on ECHOs is often performed as well, i.e.\,to detect the abnormalities in the shape of the septal wall and changes in the right-ventricular area \cite{Gali2015,Zhang2018}. 
However, the aforementioned procedures for PH estimation are time-consuming and expertise-demanding, which   % and subject to inter-observer variability.
may delay care to a more advanced stage of illness, potentially decreasing the chance of survival \cite{Barst2012}. 

\subsubsection{Related Work.} Although several machine learning methods have been proposed to automatically estimate PH in adults using different input modalities, such as chest X-rays \cite{Zou2020,Kusunose2020}, ECGs \cite{Kwon2020,Mori2021}, heart sounds recorded by sensors \cite{Kaddoura2016}, CTs \cite{Vainio2021}, and MRIs \cite{Dawes2017,Bello2019}, not much effort has been directed towards the automatic assessment of PH using echocardiography. 
The only exception is the work of Zhang et al. \cite{Zhang2018}. % where the authors show the potential of using deep learning for predicting PH using ECHOs. 
This method, however, focuses on PH in adults and has several limitations. First of all, it works on static frames of the ECHO videos and does not exploit the spatio-temporal patterns in the sequence. 
Second, it only uses a single view of the heart (apical 4-chamber view), although the literature has shown that considering multiple views improves accuracy for the manual assessment of PH \cite{Schneider2017}. 
Moreover, previous works on automatic PH detection focus on binary classification, which limits the clinical usability, because, the determination of the severity of PH from ECHOs is crucial for correct treatment, but is a challenge for cardiologists \cite{Dasgupta2021,Fisher2009}. % , with only around 47\% agreement with right heart catherisation in younger children \cite{Dasgupta2021}.
%The determination of the severity of PH  is crucial for correct treatment, as guidelines for treatment of PH depend on the severity class. 
%Furthermore, morbidity rate is significantly increased for increased severity of PH, and thus the severity of PH determines the urgency of a treatment \cite{Corris2014,Gali2015}.
%The estimation of the severity of PH from ECHOs is however a challenge for cardiologists, with around 50% less agreement to true PH severity, compared to agreement for PH detection (agreement of severity:  47% vs. 79% for detection only) \cite{Dasgupta2021,Fisher2009}.
Finally, the applicability of deep learning methods, such as \cite{Zhang2018}, is still limited in the medical domain due to their black-box nature that makes their internal mechanisms and their results opaque.

\subsubsection{Contribution.} In this work, we propose a robust and interpretable deep learning approach -- \emph{\underline{i}ntepretable \underline{p}rediction of \underline{p}ulmonary \underline{h}ypertension in \underline{n}ewborns} (IP-PHN) to predict and classify the severity of PH by utilising spatio-temporal patterns of the ECHOs from multiple views. 
To the best of our knowledge, this is the first work on multi-view video-based automated assessment of PH in newborns. 
To increase its clinical usability, we complement our predictions with saliency maps that highlight how the learned model focuses on clinically relevant cardiac structures. We show that these learned localization maps align with how clinicians subjectively assess PH.
To ensure the reproducibility of this work, the code of IP-PHN was made publicly available under \url{https://anonymous.4open.science/r/echo_classification-DE4E/}.

\section{Materials and Methods}
\subsection{Dataset}
\label{dataset}
The dataset in this work consists of 2D transthoracic echocardiography videos (ECHOs) in 5 different standard views. Retrospectively, 194 newborns were examined and 536 ECHOs were performed in a single centre by a pediatric cardiologist % from the Hospital Barmherzige Brüder Regensburg
between the years 2019-2020. All ECHOs were performed with the GE Logic S8 ultrasound machine using the transducer S4-10. %c at 6 MHz. 
A single ECHO contains a sequence of ultrasound images of the patient's heart at a specific view. The five views include a parasternal long-axis view (PLAX), apical four-chamber view (A4C), and three parasternal short-axis views; at the level of papillary muscles (PSAX-P), at the level of semilunar valves (PSAX-S), and on the apical short-axis view (PSAX-A). 
The ECHOs operate on 25 fps, and the average video length is 5 seconds, whereby each ECHO consists of around 10 heartbeats. % The % age of the cohort varies between 0 to 5.5 years and the
% median and mean age of the cohort are 8.5 and 56 days, respectively. 

The ground-truth for each ECHO was manually annotated by a pediatric cardiologist based on the visual evaluation. % stating none, mild or moderate to severe PH.
The ground truth labels differentiate between, none (65\%), mild (16\%), and moderate to severe (19\%) PH. 
Furthermore, for each ECHO its corresponding view is also annotated. 
A detailed overview of the data is provided in Supplementary Material.
The study was approved by the local ethics committee. In addition, all data is pseudonymized.

\subsection{Preprocessing and Data Augmentation}
As the first step in pre-processing the available data, we crop and mask the ECHOs to eliminate information (such as additional text or signals) outside the scanning sector and resize them to $224$x$224$ pixels using bilinear interpolation.
We then apply histogram equalization to distribute the pixel intensities to the full range of gray-scale values and normalize them.

During training, two types of image transformations are applied: intensity transformations so that the learned model is invariant to intensity variations, and spatial transformations to increase resilience against different zoom settings of the US machine and/or placements of the transducer. In particular, we apply the following random transformations to each sequence: sharpness and brightness adjustment, gamma correction, addition of salt and pepper or Gaussian noise, variation of the background with different amounts of speckle noise, rotation up to $15^\circ$, translation up to 0.1x, scaling down to 0.8x, zooming  up to 1.2x.

\subsection{Proposed Method}
\label{method}
We introduce an end-to-end deep learning approach, IP-PHN, to automatically assess PH severity without the need for tedious manual measurements of the ventricles used in standard clinical workflow. 
%The dataset is inherently imbalanced as mild and severe cases are rare compared to the healthy cases, increasing the complexity of the problem.
We assume that for each patient we have access to multiple ECHOs showing the heart from different views. % , which should contribute to the patient-level prediction. Even though the PSAX view is commonly used for subjective assessment of PH in newborns, recent work have shown that different views, such as the A4C, could also be used to predict PH \cite{EL-Khuffash2014}.
Even though a single view is commonly used for subjective assessment of PH, recent works have shown that using multiple views is beneficial for the assessment \cite{EL-Khuffash2014,Schneider2017}.
%the dataset is inherently imbalanced as mild and severe cases are rare compared to the healthy cases. Hence, we increase the robustness of the model by employing a multi-view approach. 

\begin{figure}[t!]
    \centering
    \includegraphics[width=0.95\linewidth]{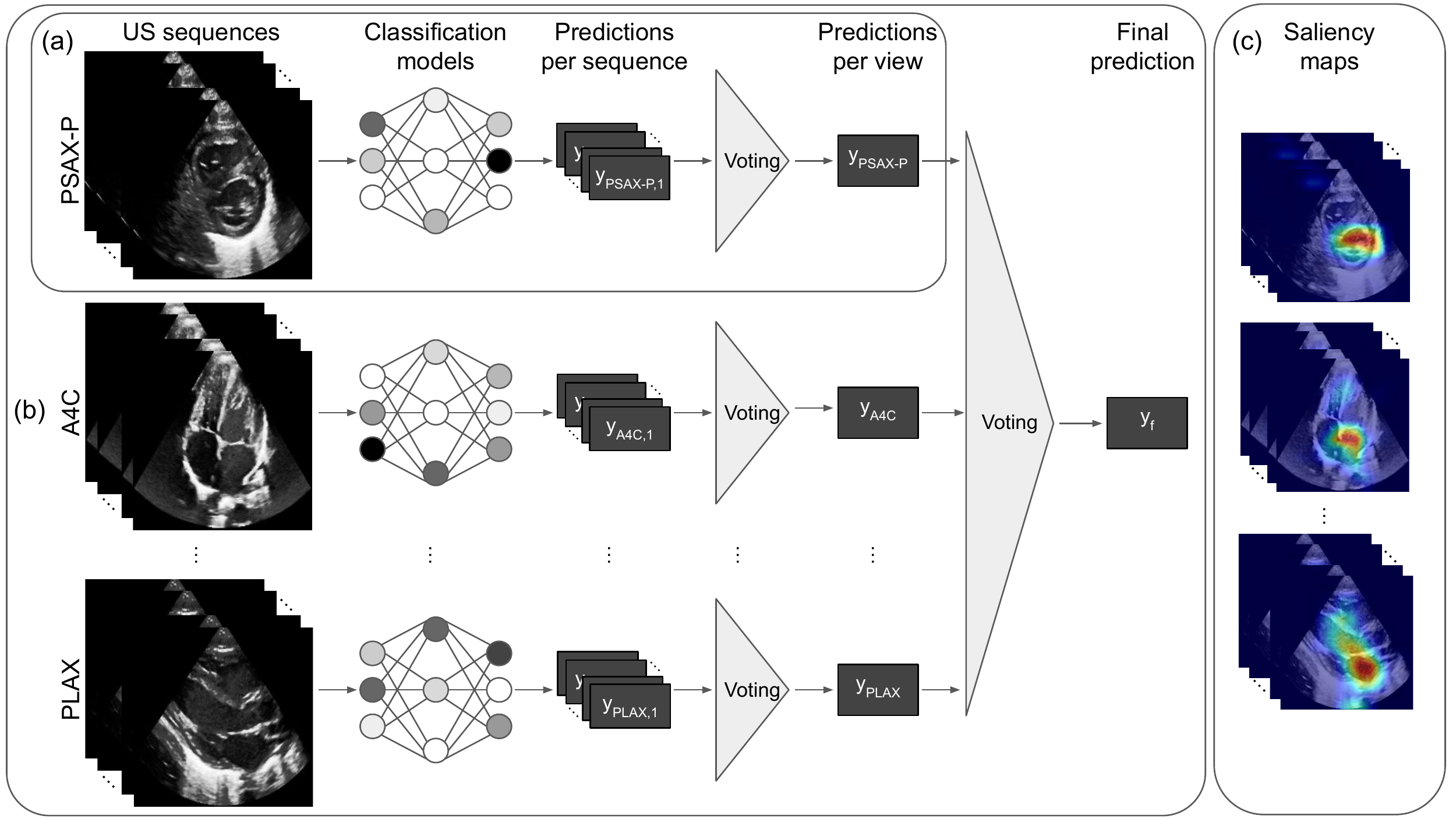}
    \caption{Overview of our proposed method, IP-PHN, to automatically assess PH severity of a patient using (a) single view and (b) multi-view approach with majority voting utilising spatio-temporal patterns of ECHOs. Spatio-temporal saliency maps (c) are provided from each view to increase clinical usability.} \label{fig:pipeline}
\end{figure}

The proposed framework is depicted in Fig.~\ref{fig:pipeline}.
% and has two key components. First, w
% \noindent \textbf{Single-view.} 
We first process each view separately. In particular, we employ a 3D-CNN architecture with residual connections and spatio-temporal convolutions across frames \cite{Hara2017} using a ResNet-18 as the model backbone \cite{Carreira2017} (see Fig.~\ref{fig:pipeline}(a)). 
In contrast to previous work \cite{Zhang2018}, our approach integrates spatial as well as temporal information into the learning process. This mitigates the frame-level variations that can occur due to external changes, such as the position or the contact of the transducer, or in the cardiac function itself. To overcome the scarcity of the annotated data, common in the medical domain, from each ECHO we extract $n$ shorter video sequences by randomly choosing a frame as their starting frame followed by $k-1$ consecutive frames, with total $k$ frames, covering on average one heartbeat. We then aggregate sequence-level predictions $\{y_{\text{view}, i}\}_{i=1,\ldots, n}$ through majority voting, i.e. by selecting the most frequently predicted label, to a view-level prediction $y_{\text{view}}$.
The view-level confidence is then defined as $C = |y^{*}_{\text{view}}| / n$, where $|y^{*}_{\text{view}}|$ is the count of the most frequently predicted label from the list of predictions for the $n$ sequences of a given ECHO per view. % , and  is the number of sequences per view.

To increase the robustness of the method further we employ a multi-view approach by combining the models trained on each available view (as in Fig.~\ref{fig:pipeline}(b)). 
The final subject-level prediction, $y_{\text{f}}$, is then achieved by majority voting of the view-level predictions. In the case of a tie, the prediction of the model(s) with higher confidence is selected. We tried different approaches for view aggregation, but simple majority voting performed the best. Note that our proposed method uses view annotations to differentiate the distinct modalities. Following recent work on view classification \cite{Zhang2018}, IP-PHN can easily be extended to incorporate ECHOs without annotation.

\subsubsection{Interpretability.} To increase the accountability and clinical usability of our proposed method, we complement our predictions with spatio-temporal saliency maps from each view (as in Fig.~\ref{fig:pipeline}(c)). The automatic localisation of relevant pixels in the video sequence for the model's prediction provides an interpretable explanation that mimics the clinical workflow. Among different methods \cite{Springenberg2015,Zhou2016,Selvaraju2017}, we chose to use Grad-CAM (Gradient-weighted Class Activation Mapping), which exploits the gradients of any target concept flowing into a given convolutional layer to produce a coarse localization map highlighting the important regions in the image for predicting the label \cite{Selvaraju2017}. Recent works \cite{Kindermans2019,Adebayo2018} have shown how pixel-space gradient visualizations, such as Guided Backpropagation and Guided Grad-CAM, could be rather insensitive to model and data, making them similar to simple edge detectors.
Grad-CAM is one of few saliency methods that pass the insensitivity check, making it our saliency method of choice \cite{Adebayo2018}.

However, Grad-CAM was originally proposed for 2D-CNNs trained on images.
We, therefore, extend Grad-CAM to 3D-CNNs processing spatio-temporal video inputs. This allows us to identify the spatio-temporal regions on the video sequence that the network finds most informative for its prediction, which are the regions in both spatial and time domains.

\section{Experiments and Results}

\subsection{Implementation Details}
Our method was developed using the Python programming language with the PyTorch deep learning library. 
Experiments were run on a cluster containing different NVIDIA GeForce graphic cards: GTX 1080, GTX 1080 Ti, RTX 2080 Ti with 2048 MB RAM per processor core. 

 %where this probability was set empirically.
For each view, we extracted $n=10$ video sequences and each sequence was composed of $k=12$ consecutive frames, covering on average one heartbeat in every sequence. To deal with class imbalance we employed a weighted random sampler, which samples elements from the dataset using their inverse class weight as their sample weight, ensuring that for every epoch the model sees approximately equal number of samples from each class. 
Additionally, during training, we continuously augmented each sample with a probability of 90\%.
%We employ a batch size of $8$ video sequences and a learning rate of $0.001$. 
Each model was trained for around $150$ epochs per view minimizing the (categorical) cross-entropy loss with the Adam \cite{kingma2014adam} optimiser. Both the learning rate and weight decay were set to $0.001$, while the batch size was set to $8$ video sequences.%, which takes around $11$ hours to train. 
%We train with cross-entropy loss using the Adam optimiser with weight decay of $0.001$.
% \begin{itemize}
%     \item Software, framework version used
%     \item Making the code publicly available, making the models publicly available?
%     \item data augmentations, size of input images
%     \item based on initial empirical tests, we employed a batch size of xx images, a learning rate of xx, and trained for xx epochs with the xx optimizer, Loss term, weight decay, drop-out.
%     \item Hyperparameter finetuning
%     \item Sensitivity regarding parameter changes
%     \item 10-fold cross validation with details on train/val split, with patients in the split and not sequences. 
%     \item Definition of the specific evaluation metrics and/or statistics used to report results
%     \item description of the computing infrastructure used (hardware and software), e.g. computation power (GPU details, RAM etc) and
%     \item average runtime for each model and result 
%     \item the results are reported on patient level. 
% \end{itemize}

\subsection{Experimental Setup}
For all experiments, a stratified 10-fold cross-validation was performed, % with replacement was performed, % (i.e. out-of-bootstrap estimate), 
such that the data was randomly split 10 times into 20\% validation set and 80\% training set. Note that, the splitting into training and validation sets was done on a patient basis. As classification metrics, we evaluated the area under the receiver operation characteristic (AUROC, one-vs-one), balanced accuracy, weighted F1-score, weighted precision, and weighted recall. The multi-view AUROC was computed from the output probabilities of the most confident model selected by the majority voting. We also report the average confidence of the models in the Supplementary Material. 
Results were averaged over the folds, and the mean and standard deviation are reported. %  on patient level. 

\subsection{Results and Discussions}
We hereby provide an empirical assessment of IP-PHN for PH severity prediction. % on the dataset described in Section \ref{dataset}.
%The results are reported in Table~\ref{tab:experiment_res}(a). % and the PH severity classification in table \ref{tab:experiment_res} (b). 
We report in Table~\ref{tab:experiment_res}(a) the quantitative performance from each of the three major views (A4C, PLAX, PSAX-P) (see Fig.~\ref{fig:pipeline}(a)), as well as from the multi-view approach (see Fig.~\ref{fig:pipeline}(b)) obtained through majority voting using different combinations of views. In particular, we combined (i) the $3$ major views (MV-3) and (ii) all views for a total of $5$ different views (MV-All). 

\begin{table}[b!]
\caption{Quantitative performance of IP-PHN for (a) PH severity prediction and (b) binary PH  detection using both single ECHO views (A4C, PLAX, and, PSAX-P) and multiple views through majority voting (MV-3 and MV-All). %\textit{MV-3} refers to the combination of A4C, PLAX, and, PSAX-P. \textit{MV-All} refers to MV-3 plus PSAX-S/A. 
The best results for each task have been highlighted in \textbf{bold}.
}
\label{tab:experiment_res}
\centering
\begin{tabular}{ l  l | c | c | c | c | c }
\hline
 & Views & AUROC & F1-Score & Precision & Recall & \makecell{Balanced\\Accuracy} \\  \hline
 \multirow{5}{*}{(a)} & A4C & 0.77$\pm 0.03$ & 0.72$\pm 0.05$ & 0.75$\pm 0.05$ & 0.72$\pm 0.05$ & 0.65$\pm 0.06$ \\ \cline{2-7}
& PLAX & 0.85$\pm 0.04$ & 0.78$\pm 0.05$ & 0.82$\pm 0.06$ & 0.79$\pm 0.06$ & 0.72$\pm 0.05$ \\ \cline{2-7}
& PSAX-P & 0.85$\pm 0.04$ & 0.81$\pm 0.05$ & 0.83$\pm 0.06$ & 0.82$\pm 0.04$ & 0.73$\pm 0.06$ \\ \cline{2-7}
 & MV-3 & 0.84$\pm 0.08$ & 0.83$\pm 0.05$ & 0.86$\pm 0.04$ & 0.83$\pm 0.05$ & 0.76$\pm 0.07$ \\ \cline{2-7}
&  \textbf{MV-All} & \textbf{0.86}$\mathbf{\pm 0.09}$ & \textbf{0.84}$\mathbf{\pm 0.06}$ & \textbf{0.86}$\mathbf{\pm 0.05}$ & \textbf{0.85}$\mathbf{\pm 0.05}$ & \textbf{0.78}$\mathbf{\pm 0.07}$ \\\hline \hline
%   \multicolumn{7}{c}{}\\
%  \cline{2-7}
% \multirow{7}{*}{(b)}& Views & AUROC & F1-Score & Precision & Recall & \makecell{Balanced\\Accuracy}\\  \cline{2-7}
\multirow{5}{*}{(b)} & A4C & 0.83$\pm 0.05$ & 0.81$\pm 0.04$ & 0.84$\pm 0.03$ & 0.81$\pm 0.04$ & 0.81$\pm 0.04$ \\ \cline{2-7}
& PLAX & 0.90$\pm 0.07$& 0.86$\pm 0.09$& 0.88$\pm 0.07$& 0.86$\pm 0.09$& 0.86$\pm 0.08$\\ \cline{2-7}
& \textbf{PSAX-P} & \textbf{0.95}$\mathbf{\pm 0.04}$ & \textbf{0.92}$\mathbf{\pm 0.03}$ & \textbf{0.93}$\mathbf{\pm 0.03}$& \textbf{0.92}$\mathbf{\pm 0.03}$ & \textbf{0.94}$\mathbf{\pm 0.03}$\\ \cline{2-7}
&  MV-3 & 0.90$\pm 0.03$& 0.87$\pm 0.04$& 0.88$\pm 0.03$& 0.87$\pm 0.04$& 0.87$\pm 0.04$\\ \cline{2-7}
&  MV-All & 0.90$\pm 0.03$ & 0.89$\pm 0.02$ & 0.90$\pm 0.02 $& 0.89$\pm 0.02$ & 0.89$\pm 0.02$\\ \hline
\end{tabular}
\end{table} % without confidence

% - Something the doctors may comment: We should probably discuss that we only evaluated the US sequences and didn't use the age information. Depending on the age the situation may be expected and resolve over time. \\

Among the single-view methods, the parasternal short-axis view at the level of papillary muscles (PSAX-P) shows the best performance in PH severity prediction and achieves an F1-score of 0.81 and Balanced Accuracy of 0.73, followed by the parasternal long-axis view (PLAX).
We observe that both PSAX-P and PLAX are fairly confident in their prediction, with an average confidence of correct predictions of 0.92 and 0.90 respectively, while it decreases to 0.83 and 0.84 for wrong predictions. Although the apical four-chamber view (A4C) is one of the most commonly used views for cardiovascular disease diagnosis, our evaluation shows that it is not as discriminative as PSAX-P and PLAX, yielding an F1-score of 0.72 and Balanced Accuracy of 0.65, which are clearly lower than the other two views. 
This is also in line with the neonatal echocardiography teaching manual \cite{EL-Khuffash2014}, where it is stated that subjective assessment of PH from the A4C view in a 2D ECHO is usually only possible for moderate to severe PH cases, and quantitative evaluation is difficult. %Our prediction on the A4C view yields F1-score of 0.72 and Balanced Accuracy of 0.65, which are clearly lower than the other two views. 

The PH severity prediction problem is challenging, not only due to the hard task at hand but also because of the data imbalance. In this case, the robustness and accuracy can be increased by utilising more views. By combining results from the PSAX-P, PLAX, and A4C views using majority voting, the F1-score improves from from $0.81$ to $0.83$, while the Balanced Accuracy improves from $0.73$ to $0.76$. %  with a relative increase ratio of $2.5\%$ and $4.1\%$, respectively.
When the other two short-axis views are furthermore included, we get an F1-score of $0.84$ and a Balanced Accuracy of $0.78$.
The majority voting is not only helpful to enhance the performance, but it is also useful in case a single view has an unsatisfactory quality for a given subject, a common scenario in many real-world applications.

Moreover, we performed an additional ablation, where we simplified the problem setting to a binary classification. We then discriminated between no PH (65\% of the data) and PH, combining mild, moderate, and severe cases, in line with previous work \cite{Zhang2018}. Even though the assessment of the severity of PH is crucial for a correct treatment, as the morbidity rate significantly increases for higher degrees of PH \cite{Corris2014,Gali2015}, the clinicians
might also be interested in simply discriminating between healthy and unhealthy patients as an initial screening procedure. In such a case, the data imbalance would be less significant.
We report the binary PH detection results in Table~\ref{tab:experiment_res}(b). PSAX-P view is still the most discriminative one, which yields improved accuracy compared to the severity prediction, with an F1-score of $0.92$ and Balanced Accuracy of $0.94$. Given the substantial prediction accuracy of the PSAX-P view alone for binary PH detection, including more views to the aggregated model does not result in increased performances.

%As an alternative approach to combining various views with majority voting, we also explored an end-to-end approach joining the views in the embedding space. This improved results compared the best single view with Balanced Accuracy from 0.73 to 0.75, but did not improve compared to the majority voting. % , which has Balanced Accuracy of 0.78. %Although the confidence is higher on average for correct predictions, in some cases the models are confidently wrong, which is why we decided not to use voting weighted by confidence. 

The existing method \cite{Zhang2018} for binary PH detection in adults does not exploit the spatio-temporal patterns. Thus, as a comparison to our proposed spatio-temporal approach, we also evaluated the spatial-only approach which proved to be inferior in both, severity prediction and binary detection tasks. The results are reported in Supplementary Material and can be used as a baseline comparison. % We achieve similar results as the existing method, with an AUROC of 0.87 compared to 0.85 in \cite{Zhang2018}, when evaluating a similar task in adults.

\subsubsection{Interpretability.}
\begin{figure}[t!]
\centering
\begin{subfigure}[b]{0.53\textwidth}
    \centering
    \includegraphics[width=\textwidth]{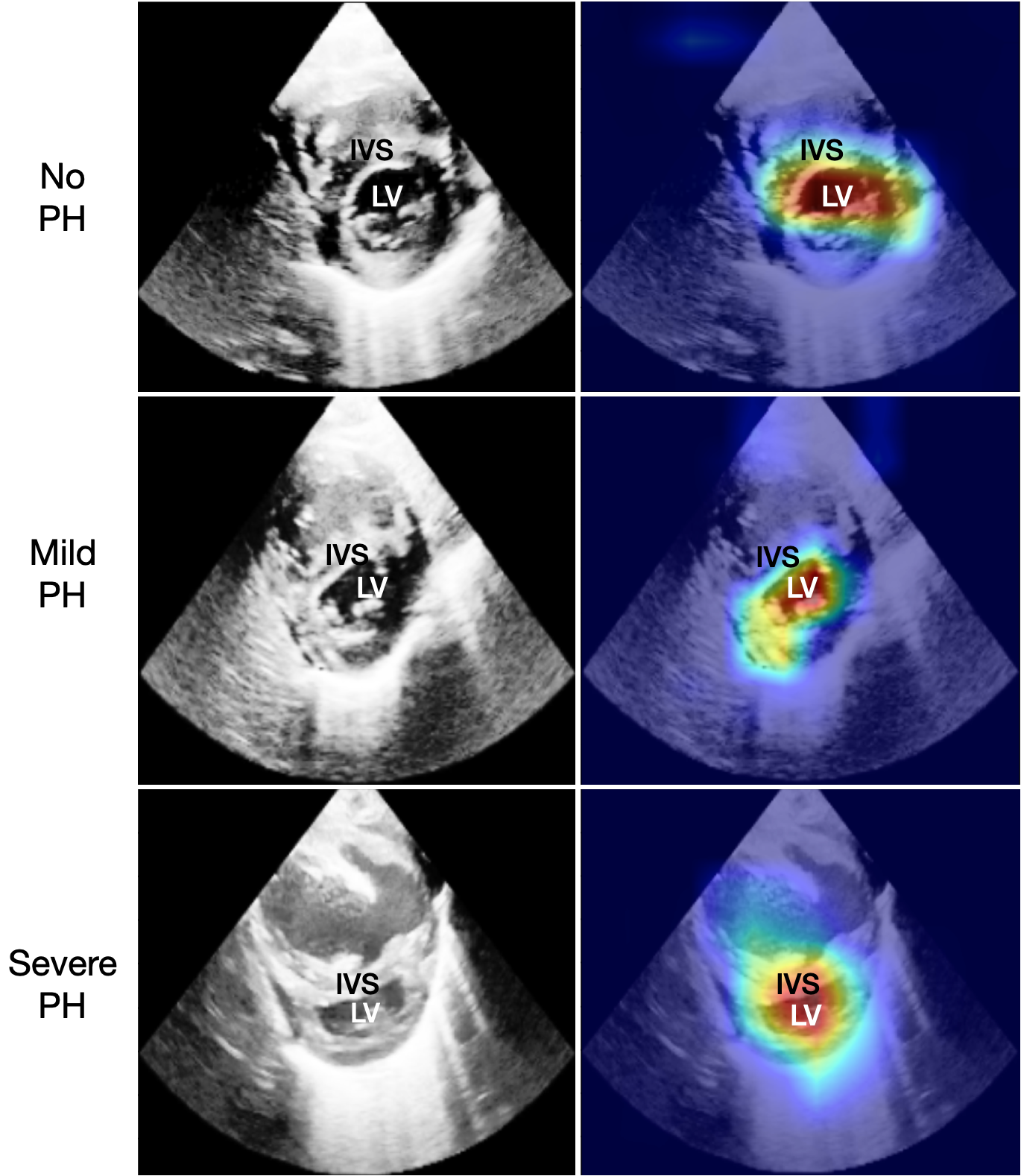}
    \caption{PSAX-P}
    \label{fig:PSAX}
\end{subfigure}
     % \hfill
\begin{subfigure}[b]{0.46\textwidth}
    \centering
    \includegraphics[width=\textwidth]{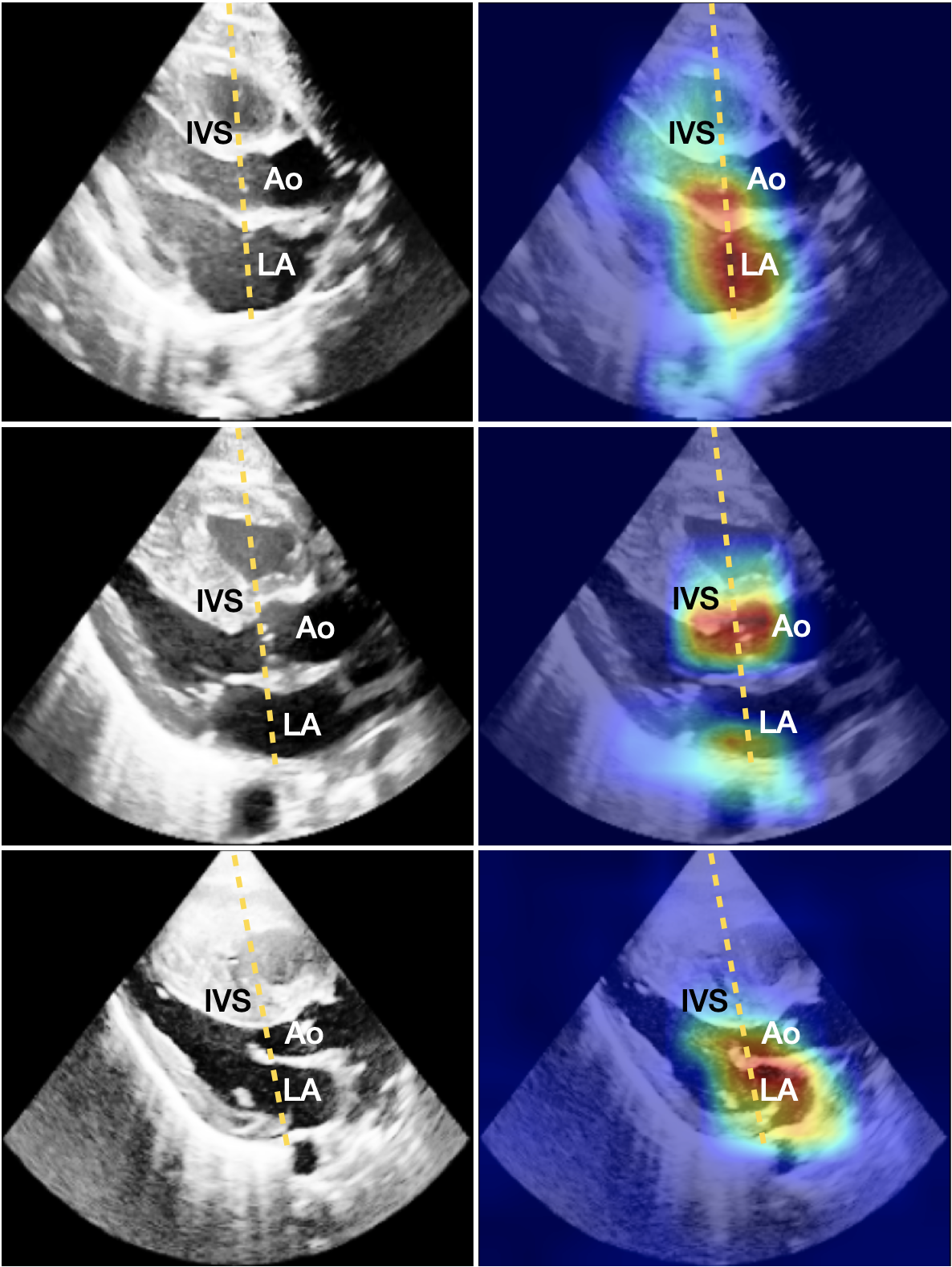}
    \caption{PLAX}
    \label{fig:PLAX}
\end{subfigure}
\caption{ECHO frames of subjects with no, mild and severe PH (left), as well as the IP-PHN saliency maps (right), for (a) the PSAX-P view, and (b) the PLAX view. The yellow line shows how the M-mode for the LA:Ao measurement is extracted. The highlighted pixels feature crucial cardiac structures.}
\label{fig:gradcam}
\end{figure}

To increase the clinical usability, our method contains a post-hoc analysis of the single-view spatio-temporal convolutions. For each ECHO view, we highlight the pixels that are the most relevant for the assessment of PH severity.
%To demonstrate the clinical usability of our method we perform a post-hoc analysis of the learned model. 
%In particular, we use Grad-CAM to highlight the pixels in the input videos that are the most relevant for PH severity prediction. 
In Fig.~\ref{fig:gradcam} we show % the resulting saliency maps (lower row) of both the (a) PSAX-P and (b) PSAL view, and we compare it with the corresponding ECHO frames for the three different level of PH (upper row).
the original ECHO frames with different levels of PH (left column) combined with saliency maps using Grad-CAM (right column) corresponding to the significant views, in (a) PSAX-P and in (b) PLAX.

According to the neonatal echocardiography teaching manual \cite{EL-Khuffash2014}, the ideal view for subjective evaluation of the intraventricular septum (IVS) morphology and left ventricle (LV) shape is PSAX-P. In mild to moderate PH the IVS becomes flat during systole and in moderate to severe PH the septum bows into the LV, such that the LV becomes D-shaped, or crescentic. We show in Fig.~\ref{fig:PSAX} that our PSAX-P severity prediction model focuses on the same clinically relevant features as are recommended for diagnosis, that is the LV and IVS.

Subjective evaluation of the IVS morphology is also possible from the PLAX view \cite{EL-Khuffash2014}. Furthermore, quantitative assessments are frequently performed on this view, including measurements of the aortic valve (AV) annulus diameter, and left atrial-to-aortic root diameter ratio (LA:Ao) by extracting the M-mode as demonstrated with the yellow line in Fig.~\ref{fig:PLAX}. When exploring the saliency map of the PLAX severity model, we see in Fig.~\ref{fig:PLAX} that the model focuses on the area around the LA, AV and Ao, and IVS. This suggests that the model is able to consider both the relevant quantitative features and the subjective ones.

Note that, for simplicity, we show the visualisation results of a single frame per patient in Fig.~\ref{fig:gradcam}. In a clinical setting, the visualizations can be viewed as a video containing spatio-temporal explanation. In Supplementary Material, we show more examples of how the focus changes along the frames of a sequence.

\section{Conclusion}
In this work we developed an automated and streamlined approach, IP-PHN, to assist clinicians in the assessment of PH in newborns. We achieved optimal performance for the severity prediction of PH when using a multi-view approach with spatio-temporal convolutional architectures using majority voting of several views.  %The severity estimation of PH from ECHOs remains a challenge for cardiologists, with around $47\%$ agreement among specialists \cite{Dasgupta2021,Fisher2009}. Given that the PH severity determines the urgency of a treatment \cite{Corris2014,Gali2015}, 
The severity estimation of PH from ECHOs is critically important as it determines the urgency of a treatment \cite{Corris2014,Gali2015}, but it remains a challenge for cardiologists \cite{Dasgupta2021,Fisher2009}. Thus,
IP-PHN may have a considerable clinical impact in increasing the accuracy and steadiness of ECHO examinations by reducing the number of late or missed diagnoses of PH. Furthermore, it may assist less trained specialists and thereby reduce the workload of highly trained experts. Finally, by highlighting the input features that are crucial for the PH assessment, the proposed approach provides interpretable explanations for the clinicians, which in turn make the system accountable.

% \subsubsection{Acknowledgements} 
\bibliographystyle{splncs04}
\bibliography{main.bib}
\newpage
\appendix
\section{Supplementary Material}

\begin{table}[h]
\caption{Characteristics of the dataset.}
\label{tab:data}
\centering
\begin{tabular}{ l | l }
\hline
Feature & Value \\ \hline
PH (\#None (\%) / \#Mild(\%) / \#Severe(\%)) & 126(65\%) / 32(16\%) / 36(19\%) \\ \hline
% Sex (\#Female (\%) / \#Male(\%)) & xx(xx\%) / xx(xx\%) \\ \hline
Age (days) (Mean $\pm$ SD) & 56 $\pm$ 160 \\ \hline
Maturity in birth (days) (Mean $\pm$ SD) & 230 $\pm$ 46 \\ \hline
Patient's weight (kg) (Mean $\pm$ SD) & 2.9 $\pm$ 1.5 \\ \hline
Manufacturer (Ultrasound Machine / Transducer) & GE Logic S8 / S4-10 at 6 MHz\\ \hline
Spatial size of original 2D images (pixels) & 1440 x 866 \\ \hline
Video length (frames) & 122 $\pm$ 2 \\ \hline
Video FPS & 25 fps \\ \hline
\end{tabular}
\end{table}
%avg len 122.45, std max len 2.7654113618049667

% Maybe just have this figure in Appendix, or not at all.
\begin{figure}[bh!]
\centering
\includegraphics[width = %\textwidth]{figs/gcam_sptmp.png}
\textwidth]{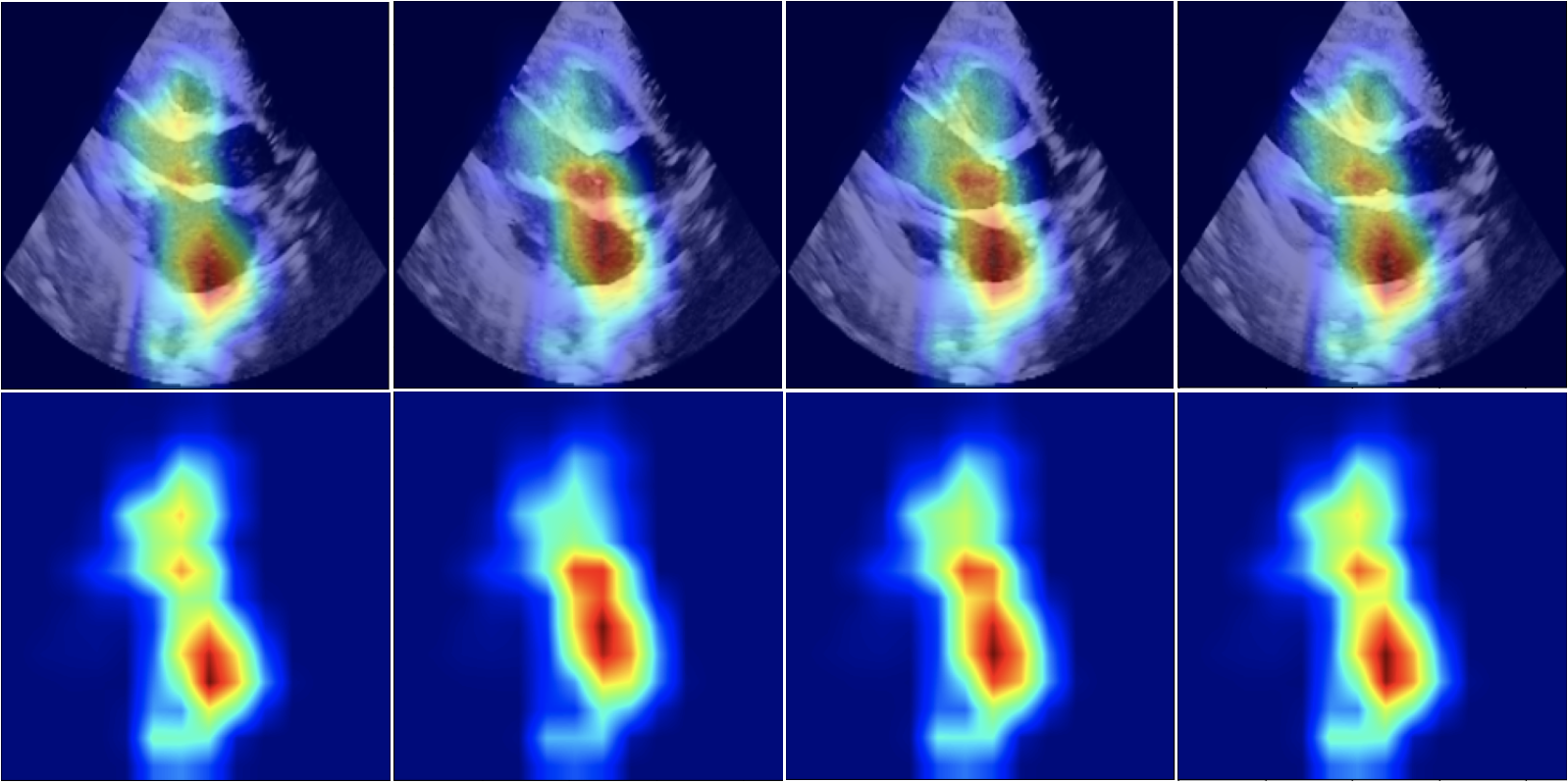}
\caption{Spatio-temporal Grad-CAM saliency maps (bottom) imposed on the original frames (top) for frames corresponding to systole, mid, diastole, mid in a PLAX ECHO.}
\label{fig:gradcam_spt}
\end{figure}

\begin{table}[th!]
\caption{Extended results from both spatial and spatio-temporal approaches for (a) PH severity prediction and (b) binary PH detection. \textit{MV-3} refers to majority voting of A4C, PLAX, and, PSAX-P views. \textit{MV-All} refers to majority voting of all five views. The best results for each task have been highlighted in \textbf{bold}. In both cases, these are spatio-temporal approaches. For the spatial approach for binary PH detection, we achieve AUROC of 0.87 on the $\textrm{A4C}^{*}$ view, compared to state-of-the-art with 0.85 when evaluating a similar task in adults.}
\label{tab:experiment_res_ext}
\centering
\begin{tabular}{ l | l | c | c | c | c | c | c}
\multicolumn{8}{c}{(a)}\\
 \hline
& View & AUROC & F1-Score & Precision & Recall & \makecell{Balanced\\Accuracy} & Confidence \\  \hline
 \multirow{7}{*}{\rotatebox[origin=c]{90}{Spatial}} & A4C & 0.79$\pm 0.04$	& 0.75$\pm 0.05$	& 0.77$\pm 0.04$	& 0.75$\pm 0.06$	& 0.67$\pm 0.06$	& 0.84$\pm 0.03$ \\ \cline{2-8}
 & PLAX & 0.84$\pm 0.04$ & 0.76$\pm 0.04$ & 0.78$\pm 0.05$ & 0.77$\pm 0.04$ & 0.70$\pm 0.05$ & 0.85$\pm 0.03$ \\ \cline{2-8}
 & PSAX-P & 0.83$\pm 0.03$ & 0.81$\pm 0.02$ & 0.82$\pm 0.03$ & 0.81$\pm 0.03$ & 0.74$\pm 0.03$ & 0.83$\pm 0.03$  \\ \cline{2-8}
 & PSAX-S & 0.74$\pm 0.07$ & 0.68$\pm 0.06$ & 0.70$\pm 0.07$ & 0.70$\pm 0.08$ & 0.62$\pm 0.06$ & 0.83$\pm 0.03$  \\ \cline{2-8}
 & PSAX-A & 0.80$\pm 0.03$ & 0.75$\pm 0.04$ & 0.76$\pm 0.04$ & 0.76$\pm 0.05$ & 0.66$\pm 0.04$ & 0.84$\pm 0.03$   \\ \cline{2-8}
 & MV-3 & 0.84 $\pm 0.03$& 0.81 $\pm 0.03$ & 0.83 $\pm 0.04$ & 0.82 $\pm 0.03$ & 0.74 $\pm 0.06$ & 0.85 $\pm 0.02$
  \\ \cline{2-8}
 & \makecell[l]{MV-All} & 0.84$\pm 0.03$ & 0.82 $\pm 0.03$ & 0.83 $\pm 0.04$ & 0.83 $\pm 0.03$ & 0.73 $\pm 0.04$ & 0.85 $\pm 0.01$
  \\ \hline \hline
 \multirow{7}{*}{\rotatebox[origin=c]{90}{Spatio-temporal}} & A4C & 0.77$\pm 0.03$ & 0.72$\pm 0.05$ & 0.75$\pm 0.05$ & 0.72$\pm 0.05$ & 0.65$\pm 0.06$ & 0.88$\pm 0.02$  \\ \cline{2-8}
 &  PLAX & 0.85$\pm 0.04$ & 0.78$\pm 0.05$ & 0.82$\pm 0.06$ & 0.79$\pm 0.06$ & 0.72$\pm 0.05$ & 0.89$\pm 0.03$ \\ \cline{2-8}
 & PSAX-P & 0.85$\pm 0.04$ & 0.81$\pm 0.05$ & 0.83$\pm 0.06$ & 0.82$\pm 0.04$ & 0.73$\pm 0.06$ & 0.90$\pm 0.03$  \\ \cline{2-8}
 & PSAX-S & 0.73$\pm 0.07$ & 0.68$\pm 0.08$ & 0.69$\pm 0.09$ & 0.69$\pm 0.08$ & 0.62$\pm 0.07$ & 0.85$\pm 0.04$  \\ \cline{2-8}
 & PSAX-A & 0.77$\pm 0.07$ & 0.74$\pm 0.06$ & 0.77$\pm 0.04$ & 0.74$\pm 0.06$ & 0.67$\pm 0.06$ & 0.84$\pm 0.04$  \\ \cline{2-8}
 & MV-3 & 0.84$\pm 0.08$ & 0.83$\pm 0.05$& 0.86$\pm 0.04$& 0.83$\pm 0.05$& 0.76$\pm 0.07$& 0.91$\pm 0.02$ \\ \cline{2-8}
 & \makecell[l]{\textbf{MV-All}} &  \textbf{0.86}$\mathbf{\pm 0.09}$  & \textbf{0.84}$\mathbf{\pm 0.06}$ & \textbf{0.86}$\mathbf{\pm 0.05}$ & \textbf{0.85}$\mathbf{\pm 0.05}$ & \textbf{0.78}$\mathbf{\pm 0.07}$ & \textbf{0.90}$\mathbf{\pm 0.02}$ \\ \hline
  \multicolumn{8}{c}{}\\
 \multicolumn{8}{c}{(b)}\\
 \hline
  & View & AUROC & F1-Score & Precision & Recall & \makecell{Balanced\\Accuracy} & Confidence \\  \hline
 \multirow{7}{*}{\rotatebox[origin=c]{90}{Spatial}} & $\textrm{A4C}^{*}$  & 0.87$\pm 0.04$ & 0.83$\pm 0.04$ & 0.85$\pm 0.03$ & 0.83$\pm 0.04$ & 0.83$\pm 0.03$ & 0.87$\pm 0.03$ \\ \cline{2-8}
 & PLAX & 0.92$\pm 0.05$ & 0.88$\pm 0.04$ & 0.89$\pm 0.04$ & 0.88$\pm 0.04$ & 0.88$\pm 0.04$ & 0.89$\pm 0.02$ \\ \cline{2-8}
 & PSAX-P & 0.93$\pm 0.04$ & 0.91$\pm 0.03$ & 0.92$\pm 0.03$ & 0.91$\pm 0.03$ & 0.92$\pm 0.03$ & 0.87$\pm 0.03$  \\ \cline{2-8}
 & PSAX-S & 0.83$\pm 0.03$ & 0.81$\pm 0.03$ & 0.83$\pm 0.02$ & 0.81$\pm 0.03$ & 0.81$\pm 0.03$ & 0.86$\pm 0.04$  \\ \cline{2-8}
 & PSAX-A & 0.86$\pm 0.04$ & 0.85$\pm 0.03$ & 0.85$\pm 0.03$ & 0.85$\pm 0.03$ & 0.84$\pm 0.03$ & 0.87$\pm 0.02$  \\ \cline{2-8}
 & MV-3 & 0.91 $\pm 0.02$ & 0.88 $\pm 0.02$ & 0.88 $\pm 0.02$ & 0.88 $\pm 0.02$ & 0.88 $\pm 0.02$ & 0.88 $\pm 0.01$ \\ \cline{2-8}
 & \makecell[l]{MV-All} & 0.92 $\pm 0.02$ & 0.90 $\pm 0.02$ & 0.91 $\pm 0.01$ & 0.90 $\pm 0.02$ & 0.90 $\pm 0.01$ & 0.87 $\pm 0.02$
  \\ \hline \hline
 \multirow{7}{*}{\rotatebox[origin=c]{90}{Spatio-temporal}} & A4C & 0.83$\pm 0.05$ & 0.81$\pm 0.04$ & 0.84$\pm 0.03$ & 0.81$\pm 0.04$ & 0.81$\pm 0.04$ & 0.91$\pm 0.03$ \\ \cline{2-8}
 & PLAX & 0.90$\pm 0.07$ & 0.86$\pm 0.09$ & 0.88$\pm 0.07$ & 0.86$\pm 0.09$ & 0.86$\pm 0.08$ & 0.91$\pm 0.02$ \\ \cline{2-8}
 & \textbf{PSAX-P} & \textbf{0.95}$\mathbf{\pm 0.04}$ & \textbf{0.92}$\mathbf{\pm 0.03}$ & \textbf{0.93}$\mathbf{\pm 0.03}$ & \textbf{0.92}$\mathbf{\pm 0.03}$  & \textbf{0.94}$\mathbf{\pm 0.03}$ & \textbf{0.90}$\mathbf{\pm 0.03}$ \\ \cline{2-8}
 & PSAX-S & 0.79$\pm 0.04$ & 0.81$\pm 0.03$ & 0.82$\pm 0.04$ & 0.81$\pm 0.03$ & 0.80$\pm 0.04$ & 0.90$\pm 0.02$ \\ \cline{2-8}
 & PSAX-A & 0.88$\pm 0.05$ & 0.87$\pm 0.03$ & 0.88$\pm 0.03$ & 0.87$\pm 0.03$ & 0.87$\pm 0.04$ & 0.89$\pm 0.03$   \\ \cline{2-8}
 & MV-3 & 0.90$\pm 0.03$  & 0.87$\pm 0.04$& 0.88$\pm 0.03$& 0.87$\pm 0.04$& 0.87$\pm 0.04$& 0.92$\pm 0.01$ \\ \cline{2-8}
 & \makecell[l]{MV-All} & 0.90$\pm 0.03$ & 0.89$\pm 0.02$& 0.90$\pm 0.02$& 0.89$\pm 0.02$& 0.89$\pm 0.02$& 0.91$\pm 0.01$ \\ \hline 
 
\end{tabular}
\end{table}

%\bibliographystyle{splncs04}
% \bibliography{ref.bib}

\end{document}